\documentclass[preprint,12pt]{elsarticle}

\usepackage[T1]{fontenc}
\usepackage[utf8]{inputenc}
\usepackage{amsmath,amssymb}
\usepackage{booktabs}
\usepackage{array}
\usepackage{tabularx}
\newcolumntype{L}{>{\raggedright\arraybackslash}X}
\newcolumntype{C}{>{\centering\arraybackslash}X}
\newcolumntype{R}{>{\raggedleft\arraybackslash}X}
\newcolumntype{P}[1]{>{\raggedright\arraybackslash}p{#1}}
\newcolumntype{Q}[1]{>{\centering\arraybackslash}p{#1}}
\newcolumntype{W}[1]{>{\raggedleft\arraybackslash}p{#1}}
\usepackage{graphicx}
\usepackage{textcomp}
\usepackage{url}
\usepackage{lineno}

\newif\ifjournalbuild
\ifdefined\JOURNAL\journalbuildtrue\fi
\newcommand{\datadoi}{10.6084/m9.figshare.35049556}

\graphicspath{{./}}

\journal{Energy and Buildings}

\begin{document}

\begin{frontmatter}

\title{A four-level executable validation protocol for evaluating LLM-generated topology
specifications of physics-based HVAC digital twins}

\author[kier,ust]{Younghwan Joo\corref{cor1}}
\ead{yhjoo@kier.re.kr}
\author[kier]{Jeonghoon Han}
\author[kier]{Sanghyun Oh}
\author[kier]{Soosik Bang}
\author[kier,ust]{Sung-il Kim\corref{cor2}}
\ead{praygod@kier.re.kr}

\affiliation[kier]{organization={Energy Efficiency Research Division, Korea
  Institute of Energy Research},
  addressline={152 Gajeong-ro, Yuseong-gu},
  city={Daejeon},
  postcode={34129},
  country={Republic of Korea}}

\affiliation[ust]{organization={Energy Engineering, University of Science \&
  Technology},
  addressline={217 Gajeong-ro, Yuseong-gu},
  city={Daejeon},
  postcode={34129},
  country={Republic of Korea}}

\cortext[cor1]{Corresponding author}
\cortext[cor2]{Corresponding author}

\begin{abstract}
Physics-based digital twins of air conditioning, a principal energy consumer in manufacturing and buildings, can be generated automatically from a declarative specification of its equipment, connections, and sizing. Large language models (LLMs) now write such specifications from plain-language descriptions. Evaluations of these specifications stop at parse success, completed runs, or textual similarity, and cannot reject a model that converges in simulation but is physically wrong. In this study, a four-level protocol is proposed that tests whether an LLM-written specification not only executes but also yields a physically valid digital twin. Each specification is executed through static validity, model build, simulation convergence, and behavioral equivalence with a hidden reference model within bands calibrated by $\pm$20\% sizing perturbations. The protocol was applied to eight LLMs and fourteen HVAC topologies, five describing an existing low-humidity test chamber. To test for authorship bias, engineers and a second model family also described six topologies. Within three repairs, seven of eight models built in 98.6 to 100\% of trials, with simulations converging in at least 80.0\%, but reached behavioral equivalence in only 50.0 to 98.6\%. About half of the failed trials converged, but not to the reference steady state, mostly in desiccant rotor systems. Pooled pass rates varied by under five percentage points across description authors. These results imply that the physics hardest to model in each field, here the desiccant rotor, must enter the generation framework as validation levels. The protocol can be adapted to other fields that generate executable digital twins from declarative specifications.

\end{abstract}


\end{frontmatter}

\ifjournalbuild\linenumbers\fi

\section{Introduction}\label{sec:introduction}

In battery manufacturing, where desiccant-wheel systems supply the dry air \cite{guan.etal2021OnsitePerformance}, keeping dry rooms at low dew points takes about 43\% of the energy of pack manufacturing \cite{yuan.etal2017ManufacturingEnergy}, and more generally, air conditioning is a principal energy consumer of manufacturing processes whose product depends on the indoor environment. Physics-based digital twins offer a way to search for better operating points of such systems, but the twin has to be built first, and construction is the recognized limiting factor: the required modeling expertise is rare at operating sites, and installations differ in equipment and topology, so a model built for one installation stops at that installation \cite{joo.etal2026AutomatedGeneration}. Automating model construction, ideally from nothing more than the descriptions and drawings a plant already has, has therefore become an active goal in industrial energy efficiency research.

Large language models (LLMs) have brought that goal within reach, and the adjacent field of building simulation has already begun adopting them. Recent studies generate EnergyPlus building energy models directly from natural-language input \cite{liu.etal2026LargeLanguage}, orchestrate multi-agent workflows that build and debug energy models \cite{zhang.etal2025AutomaticBuilding} for hundreds of buildings at a time \cite{lee.yoon2026KnowledgebasedMultiagent}, and automate the geometry-to-solver chain of urban CFD simulations from a textual prompt \cite{jiang.peng2026LargeLanguage}. In the same direction, Joo et al.~\cite{joo.etal2026AutomatedGeneration} demonstrated a framework that synthesizes an equipment-level Simulink model of a complete HVAC system from a declarative YAML topology specification, a structured document that lists the equipment of the system, its connections, and its sizing. The generated physics was validated against an operating industrial dehumidification system, and in an end-to-end demonstration a language model authored the specification from a plant description. In these studies the language model authors the model rather than assisting the modeler.

What remains unsettled is how to judge what the language model writes. Most evaluations reported so far certify that a generated model input file parses, that its entries are mutually consistent, or that a simulation run completes \cite{liu.etal2026LargeLanguage,zhang.etal2025AutomaticBuilding,jiang.chen2027LLMBEMengineerLarge}, and similarity to a reference file is still a common proxy \cite{evtikhiev.etal2023OutBLEU}. A topology specification, however, is a claim about a physical system rather than prose, and two specifications that read as near-identical text can build models whose predicted behavior differs qualitatively, for instance when a single sizing entry is off by a factor of four. Research on general-purpose code generation reached the corresponding conclusion long ago and grades programs by executing tests rather than by comparing text \cite{chen.etal2021EvaluatingLarge}. A physical model demands more than execution: the built model must also converge to a physically meaningful state, and its behavior must match the intended system within a tolerance that can be defended. A digital twin is useful only if it is physically valid in this sense, because the operating-point searches and energy predictions it is built for are computed from its converged state, and a model that builds and runs but converges to a different state does not represent the plant.

A few evaluations of building energy models and fluid networks did compare simulated behavior with a reference, but the physics through which HVAC components interact was left out. The building models had their HVAC system replaced by ideal loads \cite{jiang.etal2024EPlusLLMLarge,jiang.chen2025EfficientFinetuning}, and the fluid networks contained no HVAC equipment and were assembled from a given graph, leaving no choice of topology or sizing to the LLM \cite{sturmer.etal2026SimulationCode}. Yet this interaction is what lets a single mis-sized component shift the converged state of the whole plant. To the best of our knowledge, no evaluation has yet tested whether an LLM can choose the equipment, connections, and sizing of an HVAC plant so that the coupled heat and mass transfer among its components converges to the state of a reference model. Therefore, what is needed is an evaluation that executes a generated specification through every stage a practitioner would, grades each attempt by the last stage it passes rather than only by whether it finishes, and sets the pass criterion for equivalence so that legitimate modeling freedom is not graded as error.

In this study, a four-level executable validation protocol is proposed for evaluating LLM-generated topology specifications of physics-based HVAC digital twins. The protocol never compares a generated specification with a reference text; it executes the specification and tests not only whether it runs but also whether it yields a physically valid digital twin. Each specification is graded by the highest level reached: static validity (L1), model build (L2), simulation convergence (L3), and behavioral equivalence against a blind reference model never shown to the LLM (L4), the last judged within tolerance bands calibrated by perturbing the sizing parameters of the reference model, so that the pass criterion reflects the spread the physics permits. Error messages from every level are returned verbatim for up to three repair rounds. The protocol is demonstrated on eight LLMs spanning frontier API models, large and medium open-weight models, and a small open-weight model served on-premises. Fourteen benchmark HVAC topologies range from conventional outdoor-air handling to desiccant dehumidification with recirculation; five describe, with its equipment, the low-humidity test chamber on which the generated physics was validated. Each model-topology pair is evaluated in five repetitions. Paired control corpora written independently by human engineers and a second model family test whether description authorship affects results. Results are scored by cumulative level attainment, the share of evaluation cells (one model on one topology in one repetition) whose best attempt reaches at least each level, and every failure is recorded and classified. Finally, the demonstration is also a benchmark of how much of this generation task currently available LLMs can complete. Section 2 defines the four levels, tolerance bands, and repair loop; Section 3 describes the benchmark corpus, model roster, and evaluation procedure; and Section 4 presents and discusses the results, closing with the study limitations.

\section{The four-level executable validation protocol}\label{sec:the-four-level-executable-validation-protocol}

In this section, the generation framework is first summarized to the extent the evaluation requires; the four levels, the calibration of the tolerance bands, the repair loop, the scoring rules, and a check of passing specifications against the described operating values are then presented.

\begin{figure}[htbp]
\centering
\includegraphics[width=\textwidth]{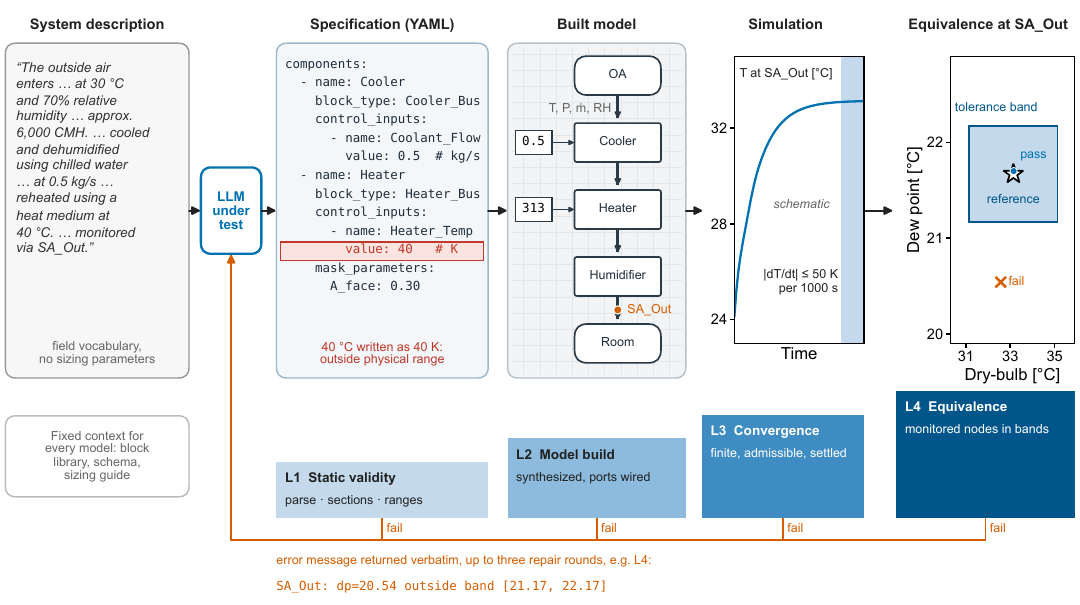}
\caption{The four-level executable validation protocol on topology 1: description excerpt, LLM-written specification (L1; the highlighted line is an illustrative unit error), built model (L2; of a specification without that error), converged simulation (L3; schematic curve), and supply-air state against its tolerance band (L4; reference state and one passing and one failing attempt of the main experiment), with the L4 error message returned for up to three repair rounds (translated from the evaluator output).}
\label{fig:protocol-overview}
\end{figure}

Figure~\ref{fig:protocol-overview} shows the protocol on the conventional outdoor-air handler (topology 1). A natural-language description in field vocabulary and field units, without any sizing parameter, is given to the model under test with a fixed context identical for all models. The returned specification is executed through the four levels in turn: static validity (L1), model build (L2), simulation convergence (L3), and behavioral equivalence (L4) at the monitored nodes within tolerance bands calibrated by perturbing the sizing parameters of a blind reference model by $\pm$20\%. When a level fails, its error message is returned verbatim for up to three repair rounds.

\subsection{Execution context: from specification to executable model}\label{sec:execution-context-from-specification-to-executable-model}

The protocol grades a specification with the generation framework of Joo et al.~\cite{joo.etal2026AutomatedGeneration}. A topology specification is a declarative YAML document that lists the components of an HVAC system, drawn from a component block library, with their connections, mask parameters, and operating inputs. From it, the framework synthesizes an executable Simulink model: every air-stream connection is a standardized psychrometric bus of temperature, pressure, mass flow rate, and relative humidity, and wiring, feedback-loop handling, and solver configuration are performed automatically. The framework and its equipment-level physics were validated against an operating industrial dehumidification system in the earlier study and are treated here as given infrastructure rather than a contribution.

The framework has two behaviors that are relevant to the grading. First, an operating input such as a coolant flow rate is declared under the component it drives with the number of the port it enters, and the framework connects it to that port unless the specification states the connection explicitly. Declaring the input in the form the framework documentation gives is therefore sufficient to obtain the intended connection, and this behavior determines which connections a built model contains, which L2 tests. Second, the heating-coil block bounds its effectiveness, the ratio of the air temperature rise to the difference between the heating-medium and inlet air temperatures, at 0.95; a larger coil therefore cannot heat the air to the temperature of its heating medium, and the sizing guide supplied with every prompt states this bound. This bound makes the perturbation envelope of some tolerance bands one-sided (Section 2.3) and limits what an oversized regeneration heater does (Section 4.4).

Each framework stage already produces a pass or fail outcome and an error message, so the four levels reuse the framework as the grading procedure, and the reference model of every topology is built by the same framework from a reference specification. No stage of the grading procedure is specific to any evaluated model.

\subsection{The four levels}\label{sec:the-four-levels}

An attempt is graded by the highest level it reaches, from level 0, which fails every check, to level 4.

Static validity (L1) checks the specification without building anything: the document must parse as YAML, contain the required sections of the schema, and state source conditions and temperature inputs inside physical ranges (temperature 200 to 500 K, pressure 80 to 120 kPa, relative humidity 0 to 1). The range check exists because unit errors are a recurring failure mode of language models in this task; a supply-air setpoint of 16\,$^{\circ}$C written as 16 K parses cleanly and fails only at this check.

Model build (L2) requires the framework to synthesize the Simulink model from the specification without errors: missing feedback handling, unwirable ports, and hallucinated component variants fail here. Because L1 does not check block types or parameter names against the component block library, any block type or parameter name the library does not contain first fails at L2.

Simulation convergence (L3) requires the built model to complete within a wall-clock limit (Section 3.3) and end in a physically admissible steady state. Every reported node value must be finite and inside the physical ranges above, and at the end of the run every node temperature must be changing by no more than 50 K per 1000 s. This threshold was set from the reference models, in which a finite simulation horizon leaves thermal capacities still approaching steady state at up to 35 K per 1000 s (Supplementary Note S1), so it passes that legitimate residual change with margin while rejecting numerical divergence. A generated model is simulated over the same horizon as its reference model and compared with it at L4 at the same instant.

For a topology with a desiccant rotor, steady state also requires the moisture balance of the rotor to close: the moisture removed from the process air and the moisture added to the regeneration air must agree within 5\%. The rotor matrix stores adsorbed moisture and takes hours of simulated time to reach a cyclic steady state; until then, the process-air outlet humidity has not reached its operating value. The temperature criterion alone does not detect this state, because node temperatures settle long before the moisture content of the rotor does. The moisture check is not applied when the rotor removes less than 0.02 g/s, as when it is switched off. The simulation horizon of each topology is that of its reference model, chosen long enough for the reference to satisfy both conditions; it is stated in the prompt, and a specification stating a different stop time is simulated over the reference horizon (Section 3.3).

Behavioral equivalence (L4) compares the generated model to a reference model at monitored nodes named in every system description by standard identifiers such as SA\_Out. Three quantities are inspected at each monitored node: dry-bulb temperature, dew-point temperature, and mass flow rate. Dew point is used instead of relative humidity because it remains resolvable in the dry-air regimes that several benchmark topologies target. The attempt passes L4 when every quantity at every monitored node falls inside its tolerance band (Section 2.3). The reference models serve as benchmark data: they are built by the framework whose generated physics was validated against field measurements in the earlier study, so equivalence to them benchmarks a generated specification against reference behavior without re-reporting any measurement.

\subsection{Perturbation-calibrated tolerance bands}\label{sec:perturbation-calibrated-tolerance-bands}

A tolerance is needed because a specification can differ from the reference without being wrong. The descriptions state operating values but withhold the sizing parameters of the components, such as tube counts and rotor diameter, apart from the given values of Section 3.1. An evaluated model therefore sizes the equipment from the sizing guide, and its choices can differ from those of the reference. A fixed percentage tolerance set by hand would reject legitimate differences if narrow and accept real errors if wide, with no physical basis for its value. The protocol instead measures the tolerance from the physics: the output spread that acceptable design variation produces is accepted as correct, and deviation beyond it is treated as a physical error.

For each reference topology, every sizing parameter stated in the reference specification was perturbed one at a time by plus and minus 20\%, a range representative of the step between adjacent standard sizes that manufacturers offer within one equipment class, and the perturbed specifications were rebuilt and simulated. A wider range would admit sizing that the guide treats as a different equipment class, and the band would excuse the sizing errors the level is meant to detect. Parameters whose perturbation does not represent design variation (absolute temperatures, initial conditions, ambient pressure, and the adsorption properties of the rotor desiccant) and parameters that the reference specification does not state were not perturbed (Supplementary Note S2). Over the perturbed variants, each monitored quantity at each monitored node spans a range from minimum to maximum, and this envelope is the tolerance band. A variant enters the envelope only if it builds and its simulation passes the convergence and physical-range checks of L3, because a diverged variant would widen the band without limit.

Mass flow rate is fixed by the source and fan specification, so its envelope can come out with essentially zero width. Temperature and dew point can also come out with zero width where no perturbed parameter moves them, as at both monitored nodes of topology 7. A zero-width band would reject a deviation of a fraction of one percent with no physical meaning. Each band was therefore given a minimum half-width, 1\% of the value for mass flow and 0.5\,$^{\circ}$C for temperature and dew point, applied in two ways that only widen the band: a band narrower than twice the minimum half-width was widened symmetrically about its center, and every band was extended where necessary to contain the reference value plus and minus the minimum half-width. The temperature quantities use an absolute minimum rather than a relative one because their values can be close to 0\,$^{\circ}$C.

The second rule matters where a physical limit stops the perturbations on one side of the reference. In a test-chamber topology whose reference reheat coil operates at the effectiveness bound of Section 2.1, a larger coil cannot raise the supply-air temperature while a smaller coil lowers it, so the envelope extends about 1\,$^{\circ}$C below the reference value and less than 0.1\,$^{\circ}$C above it; a specification 0.1\,$^{\circ}$C warmer than the reference would fail L4 while one 1\,$^{\circ}$C colder would pass. With this rule, the band contains the reference value within the minimum half-width and in this example extends from about 1\,$^{\circ}$C below the reference value to 0.5\,$^{\circ}$C above it. The limits of every tolerance band are released with the dataset, and the change of the pass rate when the width of the envelopes is halved or increased by half is reported in the last paragraph of Section 4.

\subsection{Repair loop and scoring}\label{sec:repair-loop-and-scoring}

Each evaluation cell, one model on one topology in one repetition, begins with a one-shot attempt from the standard prompt. On failure at any level, the model receives a repair prompt containing the fixed context, its previous specification, and the error message of the failing level verbatim, and submits a revised specification, up to three repair rounds. The system description is not repeated in the repair prompt. The error text is left verbatim and not summarized because the question under test is whether the grading signal itself is sufficient for recovery.

The grade of a cell is its best level: the highest level that any of its attempts reached, counting the one-shot attempt and every repair. The final attempt is recorded separately because a repair can replace a better specification with a worse one, and the share of cells whose final level falls below their best level is reported as the degradation rate. The primary measure is cumulative level attainment, the share of cells whose best level reaches at least each level. Its value at L4, the share of cells that reach L4 in any attempt up to the third repair, is called the pass rate after up to three repairs; the one-shot pass rate is the share of cells whose first attempt reaches L4. No weighted composite score is formed: any weighting would be arbitrary, and the cumulative attainment at each level shows at which level specifications fail, information that a single score would discard.

Because the L4 error message states the band interval that a monitored quantity missed, the evaluated model can, in a repair, bring a quantity into its band by changing an operating value that the description states, such as a heating-medium temperature, instead of correcting the sizing that caused the deviation. Such a specification formally passes L4 at the monitored nodes while describing a different operating condition. To measure how often this occurs, every specification that reached L4 was compared statically, without simulation, with the operating values stated in its description (Supplementary Note S3). Values that the description does not state, such as component sizes, are left to the model and are not compared. The check does not change the grade of any cell; its result is reported in Section 4.3.

\section{Experimental setup}\label{sec:experimental-setup}

Because the wording of a description depends on its author and could favor some evaluated models, the authorship of the corpus has to be controlled as well as its content. In this section, the benchmark corpus and the controls on its authorship, the model roster, and the evaluation procedure with its fairness controls are described.

\subsection{Benchmark corpus}\label{sec:benchmark-corpus}

The benchmark comprises fourteen HVAC topologies (Table~\ref{tab:setup-corpus}) spanning conventional outdoor-air handling, recirculating operation, multi-zone conditioning, and desiccant dehumidification; nine are rotor-active, meaning a desiccant rotor turns between its process and regeneration sectors and its physics participates in the closed loop. Five topologies (10, 12, 13, 14, and 16) describe the low-dew-point test chamber against which the generated physics was validated in the earlier study \cite{joo.etal2026AutomatedGeneration}, a room held at positive pressure and conditioned by a pre-cooling coil, a desiccant rotor with a regeneration heater, and an air-handling unit, at different operating points or, in topology 16, the pre-cooling coil alone. Each item is a natural-language description of the kind a field engineer would give.

\begin{table}[htbp]
\centering
\caption{The fourteen benchmark topologies, the five without an operating rotor first and then the nine rotor-active ones, each group in order of increasing number of components and, for equal numbers, of topology number. The topology numbers are the identifiers used in the released dataset and are not consecutive. Columns: number of air sources and components in the reference model, presence of a feedback loop, rotor-active status, monitored nodes, and availability of the paired control descriptions (HE, MO).}
\label{tab:setup-corpus}
\footnotesize
\setlength{\tabcolsep}{3pt}
\begin{tabularx}{\textwidth}{@{}r L Q{1.2cm} Q{1.95cm} Q{1.5cm} Q{1.1cm} P{1.65cm} Q{1.3cm}@{}}
\toprule
No. & System & Sources & Components & Feedback loop & Rotor-active & Monitored nodes & Paired controls \\
\midrule
16 & Pre-cooling coil only & 1 & 1 & no & no & PC\_Out & HE, MO \\
1 & Conventional outdoor-air handling & 1 & 4 & no & no & SA\_Out & HE, MO \\
2 & Recirculating & 1 & 4 & yes & no & SA\_Out & HE, MO \\
13 & Test chamber, rotor switched off, outdoor-air and regeneration paths closed & 1 & 4 & yes & no & SA\_Out & -- \\
3 & Multi-zone & 1 & 5 & no & no & EX\_Out & HE, MO \\
11 & Isolated rotor, design point & 2 & 2 & no & yes & DH\_Out, EA\_Out & HE, MO \\
15 & Isolated rotor, deep-dry point & 2 & 2 & no & yes & DH\_Out, EA\_Out & -- \\
4 & Desiccant with cooling & 2 & 7 & no & yes & SA\_Out, EA\_Out & -- \\
6 & Purge regeneration & 1 & 8 & yes & yes & SA\_Out, EA\_Out & -- \\
7 & Split-process regeneration & 1 & 8 & yes & yes & SA\_Out, EA\_Out & -- \\
10 & Test chamber, rotor on & 1 & 9 & yes & yes & SA\_Out, EA\_Out & -- \\
12 & Test chamber, measured flow & 1 & 9 & yes & yes & SA\_Out, EA\_Out & HE, MO \\
14 & Test chamber, deep-dry point & 1 & 9 & yes & yes & SA\_Out, EA\_Out & -- \\
5 & Desiccant with room & 1 & 10 & yes & yes & SA\_Out, EA\_Out & -- \\
\bottomrule
\end{tabularx}
\end{table}

An author received only a system diagram, a one-to-two-sentence statement of operating intent, a table of the operating values, and the writing rules, and never saw the reference specification, the component block library, or any generated model; authors with prior exposure were not assigned. The descriptions were thus authored blind, without knowledge of the reference model. Internal framework identifiers were forbidden and only field vocabulary, such as a cooling coil, was allowed. Operating values had to appear in field units ($^{\circ}$C, cubic metres per hour (CMH), \%RH), with conversion to SI units left to the evaluated model. Each description names its monitored nodes with the standard identifiers, such as SA\_Out. Sizing parameters are excluded from the descriptions: a sizing guide with representative values per equipment class is supplied identically to every model, so sizing is exercised as guide application rather than as information recall.

Values that decide the grade but cannot be stated in a description are supplied in the prompt as given values. The desiccant material of every rotor is given, because the converged state of a rotor system depends on the adsorption isotherm of the material. For the test-chamber topologies, the site values of the existing equipment are given: the geometry of the pre-cooling coil, the face and effective areas of the reheat coil, the leakage coefficient of the chamber, and, for the topology with the chamber operated as a closed recirculation loop, its volume. The given values are identical for all models and stated as fixed by the evaluator and not graded; a specification that omits or changes one is not penalized but is built with the given value.

The reference model of each topology was built by the same framework from a reference specification and simulated to the steady state of Section 2.2. For the rotor topologies, this takes 12,000 to 40,000 s of simulated time, about 3 to 11 h.

Because every grade is a comparison with a reference model, the physical consistency of the fourteen reference models was checked component by component before the experiment, from the steady-state values reported at every connection point. Two kinds of checks were made: the bounds that the second law sets (no heating coil delivers air hotter than its heating medium, no mixer delivers air outside the temperature range of its inlets, no rotor adds moisture to the process air, and no coil delivers supersaturated air) and the conservation of mass, moisture, and energy across each component. No component of any reference model violated a bound, and every rotor closed its moisture balance within 0.15\% and its air mass flows within 0.05\%. In topology 2, the room air recirculates to the mixer, so the supply-air temperature at 100 s depends on the volume and the initial temperature of the room, which the description does not state; a model built with the library defaults for these two parameters has a supply-air temperature 0.43 K below the reference, inside the tolerance band. The balances are given in Supplementary Note S1, and the component-level results are released with the dataset.

The main fourteen descriptions were authored by a single language model outside the evaluated roster (claude-sonnet-5, author code MC) under the same blind isolation. Because a corpus written by one model family could favor that family, two paired controls were added on six shared topologies: independent descriptions by practicing engineers (four authors under anonymized codes, collectively HE) and by a model from a competing family also outside the roster (gpt-5.6-terra, code MO), so that no evaluated model is graded on its own writing. The three authorships yield a paired three-way comparison on the six shared topologies (Section 4.5). The engineers returned eight descriptions, two each, which were screened for value fidelity, system fidelity, monitored-node declaration, contamination, and format. Six were used, four of them after the edits listed in Supplementary Note S4, and in two of the six the authors of this study added the chilled-water temperature of 5\,$^{\circ}$C, which the table of operating values had omitted. Two were excluded because the air path they described was incomplete or did not match the reference topology. The screening procedure and criteria are released with the dataset.

The corpus was authored in Korean to keep authorship open to engineers regardless of English fluency and translated once into English with Google Translate \cite{googleGoogleTranslate}. An author fidelity pass corrected evident mistranslations only, and every correction is logged. The frozen English versions, the corpus used in all experiments, are released with the Korean originals and the translation log.

\subsection{Model roster}\label{sec:model-roster}

Eight models were evaluated, spanning the model classes an operating site could adopt (Table~\ref{tab:setup-roster}): four frontier API models (claude-opus-5-5, claude-haiku-5-5, gpt-6.1-sol, and gemini-3.8-flash); two large open-weight models (GLM-5.3 and DeepSeek-V4-Pro) and one medium open-weight model with 27 billion parameters (Qwen3.8-27B), all three accessed through an API; and one small open-weight model with nine billion parameters (Qwen3.5-9B), served on-premises with vLLM \cite{kwon.etal2023EfficientMemory} on a workstation with four NVIDIA RTX A6000 GPUs, representing intranet-only deployments in industrial facilities. Model version strings were fixed at the start and are recorded in the table. No sampling parameters were set. Reasoning effort was set to high for every model exposing such a setting; the on-premises model ran with its server default, in which reasoning is enabled.

\begin{table}[htbp]
\centering
\caption{Model roster: model class, version string as fixed at the start of the experiment, access, and reasoning setting. All models received the identical prompt and the identical output budget of 16,384 tokens; sampling parameters were left at each provider's defaults.}
\label{tab:setup-roster}
\footnotesize
\setlength{\tabcolsep}{3pt}
\begin{tabularx}{\textwidth}{@{}P{2.85cm} P{2.1cm} L P{1.95cm} P{2.0cm}@{}}
\toprule
Model (as reported) & Class & Version string & Access & Reasoning setting \\
\midrule
claude-opus-5-5 & Frontier API & claude-opus-5-5 & API & effort high \\
claude-haiku-5-5 & Frontier API & claude-haiku-5-5 & API & effort high \\
gpt-6.1-sol & Frontier API & gpt-6.1-sol & API & reasoning effort high \\
gemini-3.8-flash & Frontier API & google/\allowbreak{}gemini-3.8-flash & API & reasoning effort high \\
GLM-5.3 & Open-weight, large & z-ai/\allowbreak{}glm-5.3 & API & reasoning effort high \\
DeepSeek-V4-Pro & Open-weight, large & deepseek/\allowbreak{}deepseek-v4-pro-20260813 & API & reasoning effort high \\
Qwen3.8-27B & Open-weight, medium & qwen/\allowbreak{}qwen3.8-27b-20260814 & API & reasoning effort high \\
Qwen3.5-9B & Open-weight, small & Qwen/\allowbreak{}Qwen3.5-9B & On-premises (vLLM, four NVIDIA RTX A6000 GPUs) & server default (reasoning enabled) \\
\bottomrule
\end{tabularx}
\end{table}

\subsection{Evaluation procedure}\label{sec:evaluation-procedure}

Every model received the same prompt for a given topology. The fixed part, identical for every model and topology, contains the component block library, specification schema, one format template, and sizing guide. The format template is a skeleton of the specification with empty fields and no sizing value; its commented examples include an outdoor-air source at 30\,$^{\circ}$C and 70\% relative humidity (Supplementary Note S5). The variable part contains the names of the monitored nodes, simulation horizon and given values of the topology, and natural-language system description. No part of the prompt is optimized for any model; the prompt totals about 15,000 to 23,000 tokens depending on the tokenizer. Five repetitions per model-topology pair give 560 cells in the main experiment; the two authorship controls add 240 cells each, for 1,040 cells in total. Each cell ran the one-shot attempt and up to three repairs (Section 2.4) under the settings of Table~\ref{tab:setup-settings}.

\begin{table}[htbp]
\centering
\caption{Evaluation settings common to all models and all corpora.}
\label{tab:setup-settings}
\small
\setlength{\tabcolsep}{4pt}
\begin{tabularx}{\textwidth}{@{}P{2.9cm} L@{}}
\toprule
Item & Setting \\
\midrule
Benchmark & 14 topologies $\times$ 8 models $\times$ 5 repetitions = 560 cells; paired controls on 6 topologies, 240 cells each (HE, MO) \\
Attempts per cell & one-shot attempt and up to 3 repairs; a cell stops at its first L4 \\
Output budget & 16,384 tokens per attempt, identical for all models \\
Sampling & provider defaults; reasoning effort set to high where the model exposes such a setting \\
Given values & values a description cannot state (desiccant material of every rotor; site equipment values for the test-chamber topologies), supplied in the prompt and enforced in the built model \\
Simulation horizon & that of the reference model: 100 s (topologies 1, 2), 300 s (16), 1,000 s (3), 6,000 s (13), 12,000 s (14), 18,000 s (11, 15), 20,000 s (4, 6, 10, 12), 40,000 s (5, 7); stated in the prompt, and a different stop time in a specification is replaced by it \\
Wall-clock limit & max(600 s, 3 $\times$ the reference wall-clock time); a run is also stopped when its simulated time advances by less than 10$^{-4}$ of the horizon within max(600 s, the reference wall-clock time / 20); both count as L3 failures \\
L3 criterion & at the end of the run, every node temperature changes by no more than 50 K per 1000 s; in rotor topologies, the moisture removed from the process air and the moisture added to the regeneration air agree within 5\% \\
L4 tolerance band & contains the reference value $\pm$ a minimum half-width (0.5\,$^{\circ}$C for temperature and dew point, 1\% for mass flow rate); beyond that, the envelope of $\pm$20\% sizing perturbations of the reference that pass L3 \\
Toolchain & MATLAB/Simulink R2024b \\
\bottomrule
\end{tabularx}
\end{table}

The output budget was 16,384 tokens for every model, and no model received a larger budget. An attempt reaching the limit is graded like any other, most often at level 0 because no complete specification was emitted, and uses up its repair round. Models were built and simulated in MATLAB/Simulink R2024b under the wall-clock limit of Table~\ref{tab:setup-settings}, which stops a simulation that cannot advance.

\subsection{Fairness controls}\label{sec:fairness-controls}

Three controls hold the comparison symmetric. First, the prompt assembly contains no model-dependent branch; the assembled prompt bytes are identical across models, and only provider API conventions differ in the transmitted request. Second, every model works under the same output budget and the same reasoning policy, with reasoning effort set to high wherever exposed. Third, the corpus authorship is controlled in both directions by out-of-roster authors from two competing families and by the human control, so that the authorship bias hypothesis is tested directly.

\section{Results and discussion}\label{sec:results-and-discussion}

In this section, the cumulative level attainment of the eight models is presented, followed by its structure across the fourteen topologies, the dynamics of the repair loop, the taxonomy of the failures, and the comparison across corpus authors. The limitations of the study are stated last.

\subsection{Cumulative level attainment}\label{sec:cumulative-level-attainment}

\begin{figure}[htbp]
\centering
\includegraphics[width=\textwidth]{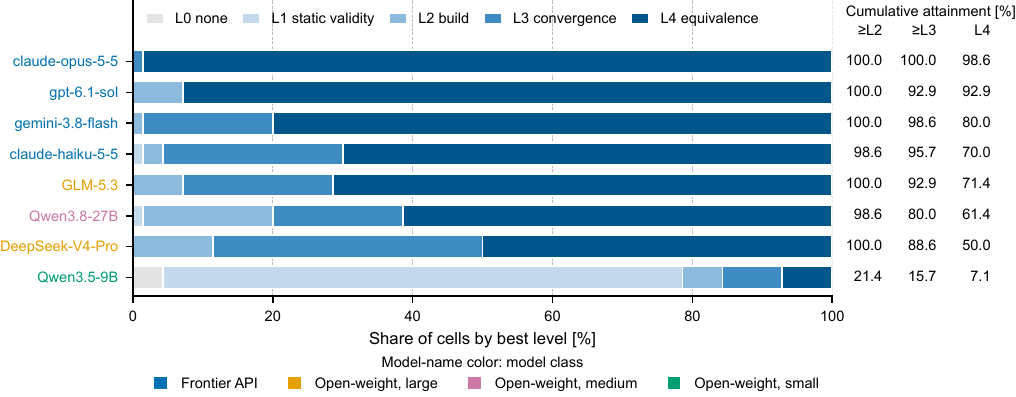}
\caption{Distribution of the 70 cells of each model over their best level in the main experiment, the highest level reached over all attempts of a cell, with the cumulative level attainment at $\geq$L2, $\geq$L3, and L4 at the right. Model names are colored by model class; one-shot results are given in Table~\ref{tab:results-pass} and Section 4.3.}
\label{fig:results-level-profile}
\end{figure}

Figure~\ref{fig:results-level-profile} shows that seven of the eight models are separated not by whether they write a buildable specification but by whether the built model reproduces the physical behavior of the reference system. Models are ordered by mean best level, with equal means ordered by the share of cells at $\geq$L3. Of the 560 cells of the main experiment, 372 (66.4\%) reached L4.

Seven of the eight evaluated models produced a buildable specification in 98.6 to 100\% of their cells, whereas the small on-premises model, Qwen3.5-9B, fell to 21.4\%. The same seven models also reached a converged simulation in most cells, from 80.0\% for Qwen3.8-27B to 100\% for claude-opus-5-5. An evaluation that stopped at build success would therefore rank seven of the eight models as equivalent, and one that stopped at convergence would separate them by at most 20 percentage points.

The models separate at the last level. claude-opus-5-5 reached L4 in 98.6\% of its cells, gpt-6.1-sol in 92.9\%, gemini-3.8-flash in 80.0\%, GLM-5.3 and claude-haiku-5-5 in 71.4 and 70.0\%, Qwen3.8-27B in 61.4\%, DeepSeek-V4-Pro in 50.0\%, and Qwen3.5-9B in 7.1\%. For five of the models, 18.6 to 38.6\% of all cells had as their best level a model that was built and converged but whose steady state lay outside the tolerance bands, the largest share being that of DeepSeek-V4-Pro (88.6\% at $\geq$L3 against 50.0\% at L4). This implies that what now separates language models on this task is whether the generated model is physically valid, that is, whether its converged state agrees with the reference system, rather than whether the specification is well formed. A model that builds and converges to the wrong state is a wrong answer rather than a partial success, and nothing in the build or convergence outcome distinguishes it from a right one.

Table~\ref{tab:results-pass} supplements the cumulative level attainment with the binary rates and the repair-loop statistics, including the total output tokens of each model. Over all models, the repair loop raised the share of passing cells from 38.2\% at one shot to 66.4\% after up to three repairs. Gains differ widely across models: gpt-6.1-sol rose from 37.1 to 92.9\% and GLM-5.3 from 25.7 to 71.4\%, whereas DeepSeek-V4-Pro rose by 5.7 percentage points, from 44.3 to 50.0\%. Budget-limited cells, those in which at least one attempt reached the 16,384-token output limit, make up 45.7\% of the claude-haiku-5-5 cells and 44.3\% of the DeepSeek-V4-Pro cells, and none of the cells of claude-opus-5-5 and gpt-6.1-sol.

\begin{table}[htbp]
\centering
\caption{Binary pass rates and repair-loop statistics by model in the main experiment (70 cells per model). The one-shot pass rate corresponds to pass@1 in the notation of Chen et al.~\cite{chen.etal2021EvaluatingLarge}.}
\label{tab:results-pass}
\footnotesize
\setlength{\tabcolsep}{3pt}
\begin{tabularx}{\textwidth}{@{}l R R R W{1.95cm} R R@{}}
\toprule
Model & One-shot pass rate {[}\%] & Pass rate after up to three repairs {[}\%] & Mean best level & Degradation rate {[}\%] & Budget-limited cells {[}\%] & Output tokens \\
\midrule
claude-opus-5-5 & 67.1 & 98.6 & 3.99 & 0.0 & 0.0 & 611,761 \\
gpt-6.1-sol & 37.1 & 92.9 & 3.86 & 1.4 & 0.0 & 560,614 \\
gemini-3.8-flash & 64.3 & 80.0 & 3.79 & 14.3 & 22.9 & 1,419,470 \\
claude-haiku-5-5 & 35.7 & 70.0 & 3.64 & 4.3 & 45.7 & 1,968,885 \\
GLM-5.3 & 25.7 & 71.4 & 3.64 & 20.0 & 27.1 & 1,346,778 \\
Qwen3.8-27B & 30.0 & 61.4 & 3.40 & 10.0 & 10.0 & 1,515,599 \\
DeepSeek-V4-Pro & 44.3 & 50.0 & 3.39 & 18.6 & 44.3 & 1,918,793 \\
Qwen3.5-9B & 1.4 & 7.1 & 1.40 & 28.6 & 18.6 & 1,516,306 \\
\bottomrule
\end{tabularx}
\end{table}

These results hold for the model versions listed in Table~\ref{tab:setup-roster} and will change as models change. Two frontier models passed more than 90\% of their cells, including 44 and 41 of their 45 rotor-active cells. The other two frontier models and the two large open-weight models passed between 50.0 and 80.0\%, with GLM-5.3 at the level of claude-haiku-5-5 (71.4 and 70.0\%); the medium open-weight model passed 61.4\%; and the small on-premises model produced a buildable specification in 21.4\% of its cells and passed 7.1\%. For a site choosing a model, build-level automation is available from every model except the smallest, whereas equivalence in more than 90\% of the rotor-active cells is at present reached by two frontier models only.

\subsection{Structure across topologies}\label{sec:structure-across-topologies}

\begin{figure}[htbp]
\centering
\includegraphics[width=\textwidth]{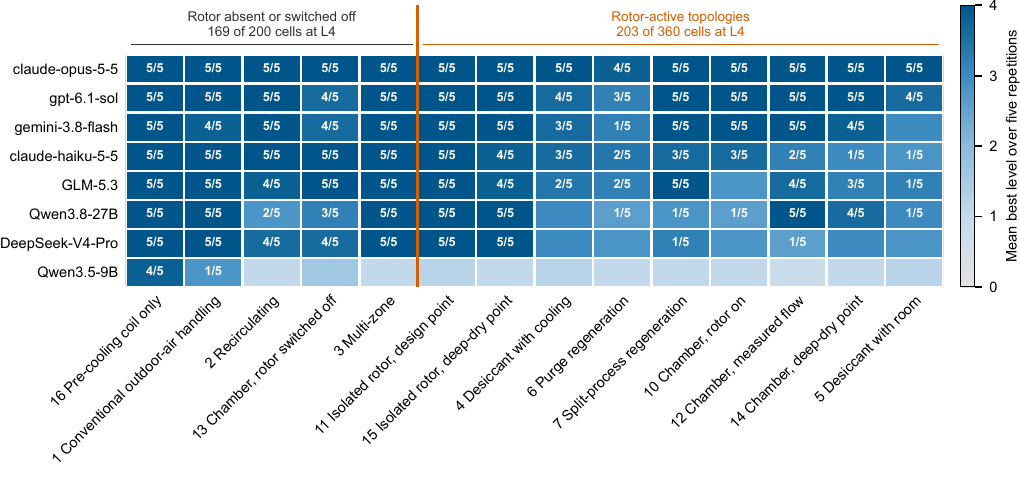}
\caption{Mean best level over five repetitions for each model and topology, with the number of repetitions reaching L4 printed where nonzero. Topologies are ordered as in Table~\ref{tab:setup-corpus}. The vertical line separates the five topologies without an operating rotor from the nine rotor-active topologies.}
\label{fig:results-matrix}
\end{figure}

Figure~\ref{fig:results-matrix} shows that the differences between the models arise in a few topologies with an operating desiccant rotor. The five topologies in which no desiccant rotor operates are four systems without a rotor and the test chamber at an operating point where its rotor is switched off and its outdoor-air and regeneration paths are closed; 169 of the 200 cells reached L4 in these five, against 203 of the 360 rotor-active cells.

The order of the models in Table~\ref{tab:results-pass} reflects a few systems rather than a uniform difference across the benchmark. The four-cell difference between claude-opus-5-5 (69 of 70 cells at L4) and gpt-6.1-sol (65) is spread over four topologies, one cell each. Six of the nine cells between gpt-6.1-sol and gemini-3.8-flash (56) come from topologies 5 and 6, in which gemini-3.8-flash passed one of its ten cells. Five of the six cells between gemini-3.8-flash and GLM-5.3 (50) come from topology 10, which GLM-5.3 did not pass in any repetition. GLM-5.3 and claude-haiku-5-5 (49) pass almost the same number of cells on different systems: claude-haiku-5-5 passed three more on topology 10 and one more each on topologies 2 and 4, and GLM-5.3 two more each on topologies 7, 12, and 14. The eight cells between Qwen3.8-27B (43) and DeepSeek-V4-Pro (35) come from topologies 12 and 14, four each, and the five passes of Qwen3.5-9B are all on topologies 16 and 1.

The rotor-active failures of the three models with the highest pass rates come almost entirely from three topologies. For regeneration air at a small flow rate that must be heated close to an 80\,$^{\circ}$C heating medium, the sizing guide includes an exception with a large regeneration heater. Topologies 4, 5, and 6 have regeneration air at about the same flow rate, and their descriptions also state a heating medium of about 80\,$^{\circ}$C, but their reference models use the regeneration heater of the default size. A specification that applies the exception to them overheats the regeneration (Section 4.4). With these three excluded, claude-opus-5-5, gpt-6.1-sol, and gemini-3.8-flash passed 100, 100, and 96.7\% of the cells of the six remaining rotor-active topologies, against 100, 96.0, and 92.0\% on the five without an operating rotor. For claude-haiku-5-5, GLM-5.3, and DeepSeek-V4-Pro the gap remains on the same comparison, 60.0, 70.0, and 40.0\% against 100, 96.0, and 92.0\%, and Qwen3.8-27B passed 70.0 and 80.0\% of the two groups.

\begin{figure}[htbp]
\centering
\includegraphics[width=\textwidth]{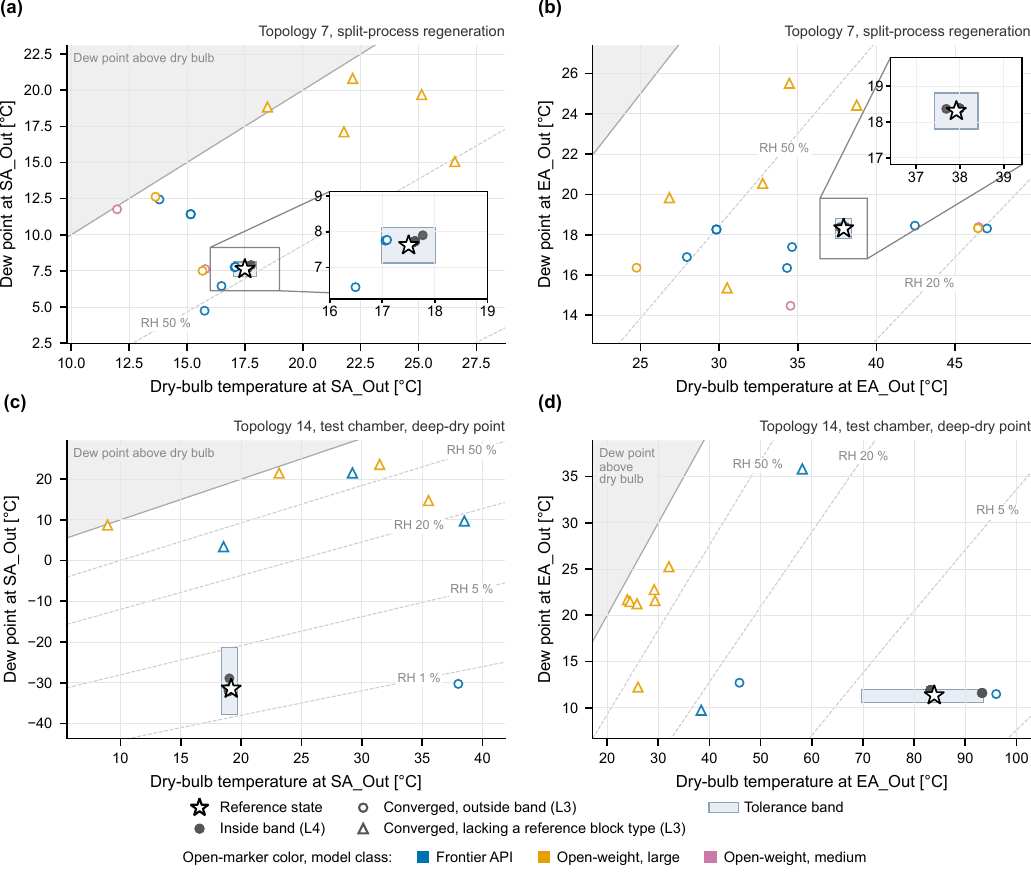}
\caption{Dew point (over liquid water) against dry-bulb temperature at SA\_Out and EA\_Out for topologies 7 (a, b) and 14 (c, d). Star: reference state (hidden from evaluated models); gray circles: attempts at L4; open markers, colored by model class: converged attempts outside the tolerance band (triangles lack a reference block type). In (a, b), three identically sized gpt-6.1-sol attempts coincide; insets enlarge $\pm$1.5\,$^{\circ}$C\@. Dashed curves: constant relative humidity; gray region: dew point above dry bulb.}
\label{fig:results-equivalence}
\end{figure}

Figure~\ref{fig:results-equivalence} shows that a specification whose simulation converged outside the band describes a different steady state of the system, not a marginally miscalibrated one. The panels show the monitored nodes SA\_Out and EA\_Out of topology 7 (split-process regeneration system) and topology 14 (test chamber at its deep-dry point). In topology 7, the 25 attempts that reached L4 lie within 0.3\,$^{\circ}$C of the reference in both dry-bulb temperature and dew point at both nodes. The 18 converged attempts outside the band lie up to 13\,$^{\circ}$C away. Each panel shows the attempts that report its node. In topology 14, two of the converged attempts outside the band contain every block type of the reference. One reaches a supply temperature 18.8\,$^{\circ}$C above the reference, and the other an exhaust temperature 38.1\,$^{\circ}$C below it. An open marker inside the tolerance band belongs to an attempt that failed in a quantity the panel does not show, either the mass flow rate at the same node or a state at the other node. The two panels of each topology are therefore read together.

The sensitivity behind these deviations is a property of the rotor physics rather than of the models. Adsorption and desorption are coupled through the rotating matrix, and in a closed loop the converged dry-bulb temperature and dew point of the whole system move by tens of degrees once the sizing of the rotor or of its regeneration heater, or their operating condition, leaves a narrow range. A wrongly sized rotor system therefore does not break the build and often does not break convergence, but produces a converged model of a different operating state. This implies that an evaluation of generated HVAC models has to compare converged states, because the errors that matter most in these systems are invisible to every check before that comparison.

\subsection{Repair-loop dynamics}\label{sec:repair-loop-dynamics}

\begin{figure}[htbp]
\centering
\includegraphics[width=\textwidth]{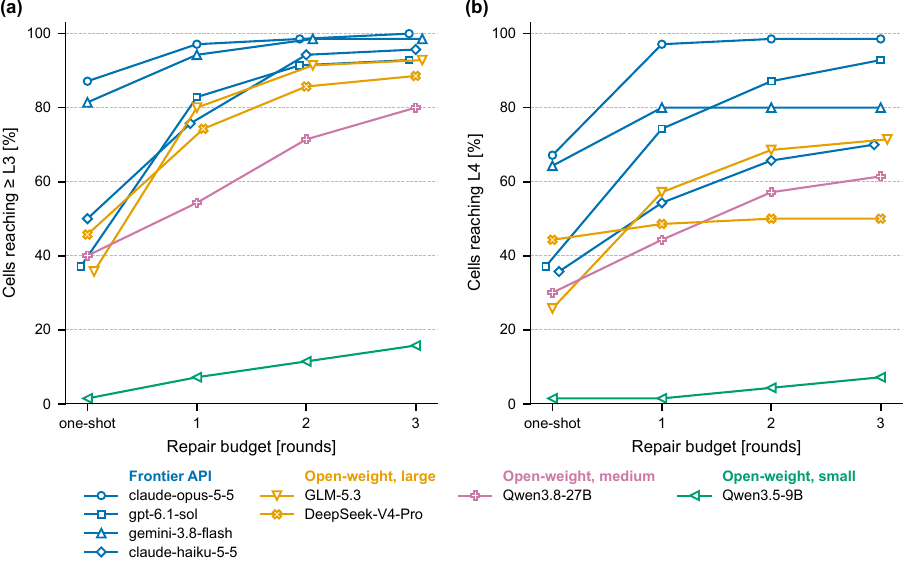}
\caption{Repair-loop dynamics in the main experiment: share of cells reaching (a) at least L3 and (b) L4 against the repair budget, by model. Marker shape identifies the model and color the model class; points of the same budget within 2 percentage points of each other are offset horizontally.}
\label{fig:results-repair}
\end{figure}

Figure~\ref{fig:results-repair} shows that most of the gain of the repair loop is made in the first repair. The L4 share of claude-opus-5-5 rises from 67.1\% at one shot to 97.1\% after one repair and by 1.4 percentage points after that; that of gemini-3.8-flash rises from 64.3 to 80.0\% after one repair and stays there; and that of DeepSeek-V4-Pro rises by 5.7 percentage points over the three repairs. The L4 share of gpt-6.1-sol rises at every round (37.1, 74.3, 87.1, and 92.9\%), and its $\geq$L3 share follows almost the same course (37.1, 82.9, 91.4, and 92.9\%). Every one-shot attempt of gpt-6.1-sol that converged also reached L4, and its other one-shot attempts failed before convergence, so its repairs moved cells past the build and convergence levels to equivalence.

Part of the flattening comes from the output limit rather than from the feedback. Of the 93 cells whose best attempt converged but failed L4, 73 contain at least one attempt cut off at the 16,384-token output limit; such an attempt is graded at level 0 in most cases and uses up a repair round. All 14 failed cells of gemini-3.8-flash and 30 of the 35 failed cells of DeepSeek-V4-Pro contain such an attempt, against none of the six failed cells of claude-opus-5-5 and gpt-6.1-sol.

\begin{figure}[htbp]
\centering
\includegraphics[width=\textwidth]{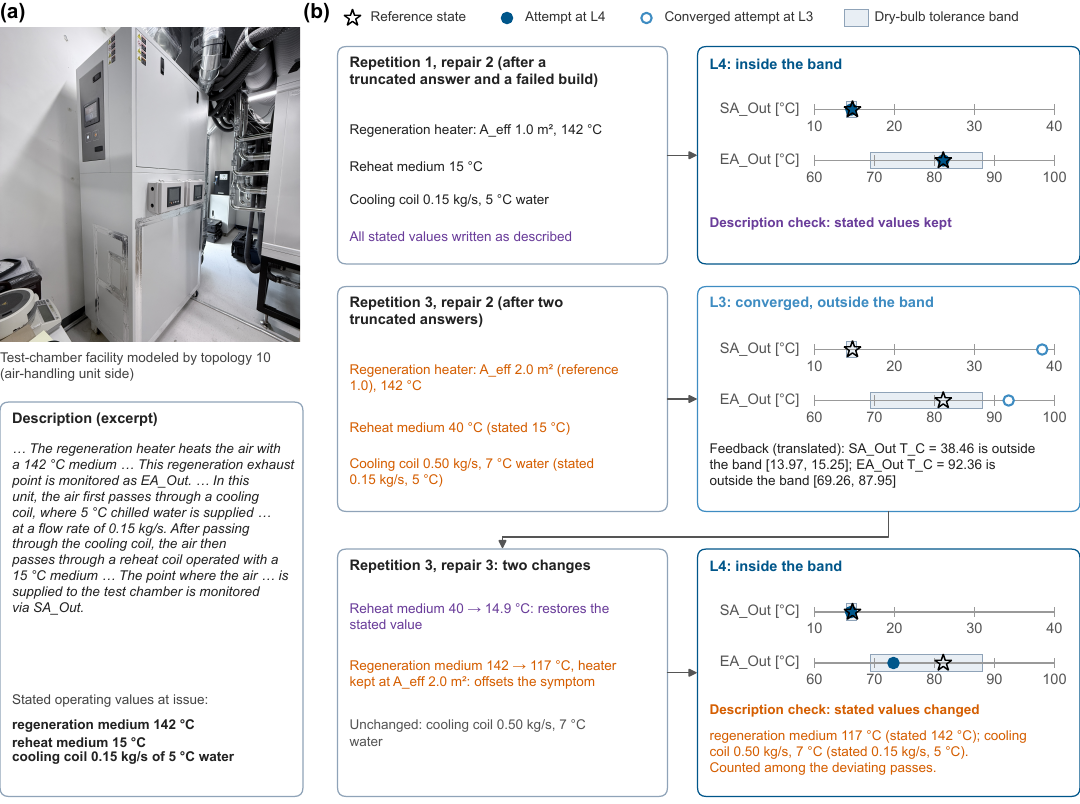}
\caption{Comparison of two repair paths of claude-haiku-5-5 on topology 10: (a) the test-chamber air-handling unit and the stated operating values; (b) specifications, levels, and the description check of repetitions 1 and 3. In (b), the open star is the reference state, a filled circle an attempt at L4, an open circle a converged attempt at L3, and the shaded bar the tolerance band in dry-bulb temperature; a filled circle that coincides with the reference lies under the star.}
\label{fig:results-decision-trace}
\end{figure}

Because the L4 error message states the band interval that a monitored quantity missed, a cell at L3 can reach L4 in two ways: by correcting the cause of the deviation, or by changing an operating value of the description until the monitored quantity falls inside its band. Of the 372 cells of the main experiment that reached L4, 13 did so with a specification whose operating values differ from those of the description (nine of claude-haiku-5-5, two of gpt-6.1-sol, and one each of Qwen3.8-27B and Qwen3.5-9B); in 11 of the 13, the passing attempt followed an L4 error message. Figure~\ref{fig:results-decision-trace} follows two cells of claude-haiku-5-5 on topology 10 that start from the same description and both reach L4, one with the operating values of the description, which a specification must keep, and one after changing them. In repetition 1, the one-shot answer was cut off at the output limit and the first repair failed to build; the second repair kept the operating values of the description and a regeneration heater of the reference size, and reached L4 with a regeneration exhaust temperature of 81.5\,$^{\circ}$C, the same as the reference to within 0.1\,$^{\circ}$C\@. In repetition 3, the second repair converged with a reheat medium at 40\,$^{\circ}$C instead of the described 15\,$^{\circ}$C, a cooling-coil water flow of 0.5 kg/s at 7\,$^{\circ}$C instead of 0.15 kg/s at 5\,$^{\circ}$C, and a regeneration heater with twice the reference effective area, which put the supply air at 38.5\,$^{\circ}$C and the regeneration exhaust at 92.4\,$^{\circ}$C, both outside their bands. The third repair restored the reheat medium to 14.9\,$^{\circ}$C, which brought the supply air back into its band, and lowered the regeneration medium from the described 142\,$^{\circ}$C to 117\,$^{\circ}$C while keeping the oversized heater, which brought the exhaust to 73.2\,$^{\circ}$C, inside its band. The cell passed L4, and the check against the described operating values (Section 2.4) counts it among the deviating passes, because the regeneration medium and the cooling-coil values differ from the description.

\begin{figure}[htbp]
\centering
\includegraphics[width=\textwidth]{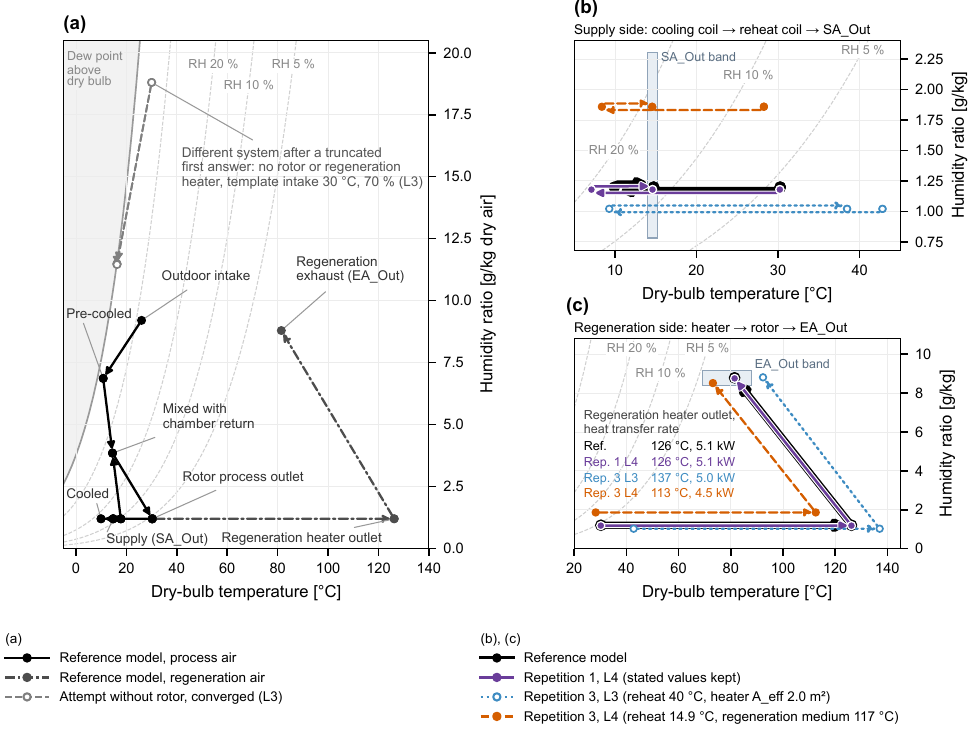}
\caption{Psychrometric air paths of topology 10: (a) reference model (process air solid black, regeneration air dash-dotted dark gray) and a converged DeepSeek-V4-Pro attempt without the rotor (gray dashed); (b) supply and (c) regeneration sides of repetitions 1 and 3 of claude-haiku-5-5, with tolerance bands. Arrows follow the air flow; in (b) the cooling and reheat segments, of equal humidity ratio, are drawn slightly apart. Dashed curves: constant relative humidity; gray region in (a): dew point above dry bulb.}
\label{fig:results-psychrometric}
\end{figure}

Figure~\ref{fig:results-psychrometric} shows the same two cells on the psychrometric chart, where the two passes differ at the monitored nodes, inside the tolerance bands, and by more along the air path. In panel (b), repetition 1 coincides with the reference at SA\_Out and the second repair of repetition 3, which converged without reaching L4, ends 24\,$^{\circ}$C warmer, outside the band. At SA\_Out, where the reference dew point is $-$14.8\,$^{\circ}$C and the band extends from $-$19.9 to $-$6.6\,$^{\circ}$C, the passing attempt of repetition 1 has a dew point of $-$15.0\,$^{\circ}$C and that of repetition 3 $-$9.4\,$^{\circ}$C, both inside the band. In panel (c), the regeneration heater of repetition 1 heats the air to 126\,$^{\circ}$C with a heat transfer rate of 5.1 kW, as in the reference, whereas that of the passing attempt of repetition 3 heats it to 113\,$^{\circ}$C with 4.5 kW, and its exhaust lies 8\,$^{\circ}$C below the reference value but inside the band. This implies that agreement at the monitored nodes admits a model whose unmonitored states, here the regeneration air leaving the heater, differ from the reference by 13\,$^{\circ}$C when an operating value of the description has been changed, which is the case that the check against the described operating values counts.

The repair prompt also does not repeat the system description (Section 2.4), and some repairs replaced the described system with a simpler one. The gray dashed path in panel (a) of Fig.~\ref{fig:results-psychrometric} is such an attempt of DeepSeek-V4-Pro on topology 10: its one-shot answer was cut off at the output limit, and its second repair built a system without the rotor and the regeneration heater from an outdoor intake at 30\,$^{\circ}$C and 70\% relative humidity, the values of the format template rather than those of the description; all eleven converged attempts that replaced the system on this topology used the same intake. Counting only converged attempts that contain every block type of the reference specification, the $\geq$L3 share of DeepSeek-V4-Pro falls from 88.6 to 54.3\%, and that of all models from 83.0 to 77.5\%. None of the repair attempts that replaced the system reached L4. The one attempt that reached L4 without a block type of the reference is a one-shot specification of claude-haiku-5-5 for the conventional outdoor-air handler that omits the room downstream of the monitored supply node and is counted among the deviating passes because the room loads of the description are missing.

\subsection{Failure taxonomy}\label{sec:failure-taxonomy}

\begin{figure}[htbp]
\centering
\includegraphics[width=\linewidth]{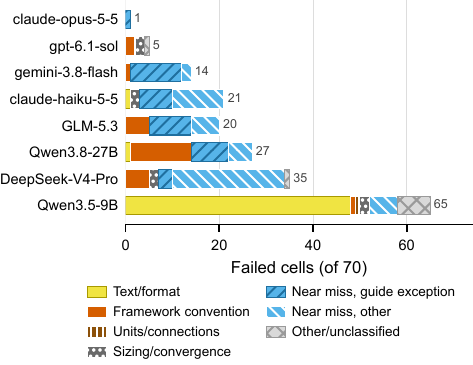}
\caption{Classification of the 188 main-experiment cells that did not reach L4 by the error of their best attempt, grouped by the field in which the error originates, with counts by model.}
\label{fig:results-taxonomy}
\end{figure}

Figure~\ref{fig:results-taxonomy} shows that about half of the failures are specifications that were built and converged but whose steady state disagreed with the reference. The figure classifies the 188 cells of the main experiment that did not reach L4 by the error of their best attempt, grouped by the field in which the error originates: the text the model writes (text and format), the conventions of the modeling framework (here, the declaration of recirculation loops), physical units and connections, physical sizing and convergence, and near misses, the cells whose best attempt failed only at L4; 9 cells have other or unclassified errors. Near misses account for 93 of the 188 cells (49.5\%).

The errors that arise outside the physics are concentrated in two models. Of the 50 cells that failed on text and format (invalid model names, block or parameter names that do not exist in the library, invalid parameter values, and schema violations), 48 belong to Qwen3.5-9B, and of the 27 cells that failed on the framework convention, 13 belong to Qwen3.8-27B\@. The convention concerns recirculation: where air returns to an earlier point of the system, the framework breaks the loop in the model with a delay block, and the specification must declare the loop with the initial state of the returning air as an air-stream bus, which these cells omitted or wrote in another form. One cell failed on physical units (a rotor temperature parameter written in kelvin where degrees Celsius were expected), and 8 cells failed on sizing or convergence before a steady state was reached, for example because the rotor state equation had no finite derivative at the first time step.

Of the near misses, 39 cells come from one sentence of the sizing guide. For regeneration air at a small flow rate, about 0.5 kg/s, that must be heated close to an 80\,$^{\circ}$C heating medium, the guide gives an exception with a large regeneration heater (a face area of 0.15 m$^{2}$ and an effective area of 10.0 m$^{2}$). The reference model of topology 7 uses this heater, and those of topologies 4, 5, and 6 use the default one, although all four descriptions state only a heating medium of about 80\,$^{\circ}$C and not the heat transfer rate. For these systems, the description and the sizing guide therefore admit both heaters, and the reference model contains one of them. A specification that uses the large heater in topologies 4, 5, or 6 heats the regeneration air close to the effectiveness bound of Section 2.1, overheats the regeneration, and converges outside the band. The 39 cells are the near misses on topologies 4, 5, and 6 whose regeneration heater has an effective area of at least 9 m$^{2}$: 11 of gemini-3.8-flash, 9 of GLM-5.3, 8 of Qwen3.8-27B, 7 of claude-haiku-5-5, 3 of DeepSeek-V4-Pro, and 1 of claude-opus-5-5. These 39 cells differ from the reference model in a sizing choice that the prompt leaves open, and six of the eight models have at least one. Because L4 grades equivalence with the reference model, the pass rates above count these cells as failures; with them counted as passes, the pass rate becomes 73.4\% instead of 66.4\% (last paragraph of Section 4). This implies that the sizing guide is itself part of what the protocol evaluates.

Of the remaining 54 near misses, 46 are on the rotor-active topologies, and 24 belong to DeepSeek-V4-Pro and 11 to claude-haiku-5-5. The quantities they violate are those set by dehumidification and regeneration: over all 93 near misses, the exhaust temperature at EA\_Out is outside its band in 69 cells and the supply dew point at SA\_Out in 65. The field at which each model fails differs between models: Qwen3.5-9B fails mostly on text, Qwen3.8-27B on the framework convention as well as on near misses, gpt-6.1-sol in five cells (two each on the framework convention and sizing, and one unclassified), and the other five models mostly on near misses. This implies that the protocol does more than rank models, since the field of the failure states what a model lacks.

\subsection{Corpus-author comparison}\label{sec:corpus-author-comparison}

\begin{figure}[htbp]
\centering
\includegraphics[width=\linewidth]{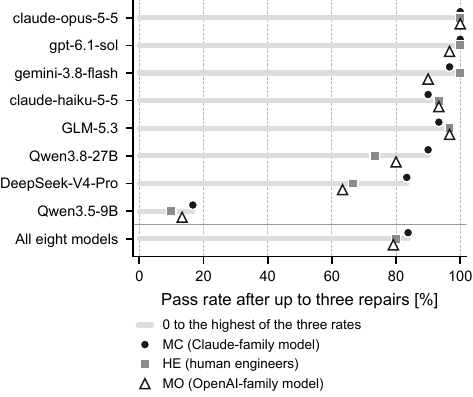}
\caption{Pass rate after up to three repairs on the six paired topologies by corpus authorship (MC, HE, MO), for each model and for all eight models pooled, 30 cells per model and 240 pooled. The gray bar extends from 0 to the highest of the three rates, and the three markers of a row are offset vertically.}
\label{fig:results-author3way}
\end{figure}

Figure~\ref{fig:results-author3way} shows that the pass rate changes with the author of the description by a few cells, and that the change is concentrated in two models. The three corpora, compared on the six topologies for which all three exist, are the main corpus written by a Claude-family model outside the roster (author code MC), the descriptions written by practicing engineers (HE), and those written by an OpenAI-family model outside the roster (MO).

Pooled over all models, 192 of 240 cells reached L4 on the engineers' descriptions (80.0\%; 95\% Wilson confidence interval 74.5 to 84.6\%), 201 on the main corpus (83.8\%; 78.6 to 87.9\%), and 190 on the second model family's descriptions (79.2\%; 73.6 to 83.8\%). Four of the nine cells by which the main corpus exceeds the engineers' descriptions come from topology 12 (test chamber with measured flow); by model, Qwen3.8-27B and DeepSeek-V4-Pro each pass five more cells on the main corpus. The second model family's descriptions pass eleven cells fewer than the main corpus, seven on topology 12.

The hypothesis under test is that the main corpus favors the models of the family that wrote it. The main corpus gives the two Claude-family models in the roster no advantage: claude-opus-5-5 passed all 30 cells on each of the three corpora, and claude-haiku-5-5 27 cells on the main corpus against 28 on each of the other two; gpt-6.1-sol passed 30 on the main corpus and on the engineers' descriptions and 29 on those of its own family. A description written by a model is not uniformly easier than one written by an engineer: the second model family's descriptions pass 190 cells against 192 for the engineers'. The remaining difference is an advantage of the main corpus for two open-weight models: four of the five cells of Qwen3.8-27B come from the chamber topology with measured flow, and those of DeepSeek-V4-Pro are spread over three topologies.

Four limitations qualify these results. First, the benchmark covers one domain, fourteen topologies, and a single prompt assembly, and transfer to another domain requires its own reference models and monitored-node contract, subject to validation there. Within the benchmark, L4 establishes behavioral equivalence with a reference model, not with measurements, and no reference model was compared with measurements in this study; the framework and its component physics were validated with measurements on the test chamber in the earlier study, five of the fourteen topologies describe that installation, and the reference models of the other nine are assembled from the same component block library. Second, the human control covers six of the fourteen topologies and four authors, so the authorship comparison is established at that scale. Third, the output and repair budgets were fixed for all models at 16,384 tokens and three repairs, and 73 of the 93 near-miss cells contain an attempt cut off at the output limit; whether a larger budget or more repairs would bring these cells to L4 was not tested and remains a subject for future work. Finally, the pass rate of 66.4\% counts every pass that the protocol defines; it becomes 75.0\% without topologies 4 to 6, 73.4\% with the guide-exception near misses counted as passes, and 64.1\% with the passes that changed a described operating value counted as failures. When the stored attempts are graded again with the envelope of every tolerance band halved in width or increased by half and the minimum half-width unchanged, the pass rate is 58.0 and 68.6\%, respectively, and the order of the three models with the highest pass rates does not change.

\section{Conclusion}\label{sec:conclusion}

In this study, a four-level executable validation protocol was proposed that tests whether an LLM-generated topology specification of an HVAC digital twin executes and yields a physically valid twin. Each specification is executed through static validity, model build, simulation convergence, and behavioral equivalence against a blind reference model within perturbation-calibrated tolerance bands. The protocol was demonstrated on eight LLMs over fourteen HVAC topologies, with control corpora from practicing engineers and a second model family. From the comparison, it was found that seven of eight models produced a buildable specification in 98.6 to 100\% of cells and a converged simulation in 80.0 to 100\%, whereas behavioral equivalence was reached in 50.0 to 98.6\% by the same seven models, and that 93 of the 188 cells that did not reach equivalence held specifications that built and converged. Across the three corpus authorships the pooled pass rate ranged from 79.2 to 83.8\%, and the two models of the family that wrote the main corpus did not pass more cells on it.

The value of the protocol lies in where a specification fails rather than in whether it passes. Grading by build success would have rated seven models as equivalent. Grading by equivalence shows that the models differ in whether the generated model is physically valid: a wrongly sized rotor system converges to a different operating state that only the equivalence level detects, and a repair that offsets a deviation by changing a described operating value is found only by checking those values. More generally, the physics hardest to model in each field, here the desiccant rotor and its regeneration, has to be built into the generation framework as a validation level, with the quantities to monitor and the band the physics permits set from domain knowledge. The demonstration is a benchmark of how much of this task the listed model versions complete as of this experiment.

Therefore, the protocol is expected to be transferable to other fields in which executable digital twins are generated automatically from declarative specifications, and usable as the inner loop of a tool-augmented agent that runs the level checks on its own draft. The gain such an agent achieves on the rotor-active systems remains a subject for future work.

\section*{Acknowledgements}

This work was conducted under the framework of the Research and Development Program of the Korea Institute of
Energy Research (KIER) (C6-2419-63).

\section*{Data availability}

The data supporting this study are openly available at figshare (DOI: \datadoi). The archived dataset comprises
the benchmark corpus, that is, the descriptions of the fourteen topologies in the frozen English version used in
all experiments and in the Korean original, with the translation log and the screening procedure; the prompts as
the evaluated models received them; the fourteen reference specifications, the reports of the reference models,
and the component-level consistency checks; every tolerance band, with the monitored values of the perturbed
variants that set it; the record of every attempt, with the specification extracted from the model response, the
level reached, and the error message returned to the model; the reports of the converged attempts; and the tables
behind the tables and figures of this paper. A data dictionary documents every field.

The full text of the model responses was not stored during the experiment; the specification extracted from each
response is archived as it was graded. The generation framework, the Simulink implementation of the component
blocks, and the generated models are not released: a patent application on the generation method is pending, and
the component implementations remain under development. The evaluation code runs on this framework and is not
released either. Consequently, levels L2 to L4 cannot be re-run from the archive, whereas the counts, rates, and
node values reported in this paper can be recomputed from the archived records, specifications, and band files.

\section*{Declaration of generative AI and AI-assisted technologies in the manuscript preparation process}

During the preparation of this work the authors used generative AI tools in order to improve the readability and
language of the manuscript. After using these tools, the authors reviewed and edited the content as needed and
take full responsibility for the content of the published article.

\bibliographystyle{elsarticle-num}
\bibliography{../manuscript/references,local_additions}

\begin{thebibliography}{10}
\expandafter\ifx\csname url\endcsname\relax
  \def\url#1{\texttt{#1}}\fi
\expandafter\ifx\csname urlprefix\endcsname\relax\def\urlprefix{URL }\fi
\expandafter\ifx\csname href\endcsname\relax
  \def\href#1#2{#2} \def\path#1{#1}\fi

\bibitem{guan.etal2021OnsitePerformance}
B.~Guan, T.~Zhang, X.~Liu, On-site performance investigation of a desiccant
  wheel deep-dehumidification system applied in lithium battery manufacturing
  plant, Energy and Buildings 232 (2021) 110659.
\newblock \href {https://doi.org/10.1016/j.enbuild.2020.110659}
  {\path{doi:10.1016/j.enbuild.2020.110659}}.

\bibitem{yuan.etal2017ManufacturingEnergy}
C.~Yuan, Y.~Deng, T.~Li, F.~Yang, Manufacturing energy analysis of lithium ion
  battery pack for electric vehicles, CIRP Annals 66~(1) (2017) 53--56.
\newblock \href {https://doi.org/10.1016/j.cirp.2017.04.109}
  {\path{doi:10.1016/j.cirp.2017.04.109}}.

\bibitem{joo.etal2026AutomatedGeneration}
Y.~Joo, J.~Han, S.~Oh, S.~Bang, S.-i. Kim, Automated generation of
  experimentally calibrated digital twins for desiccant-based low-dew-point
  air-conditioning systems from declarative topology specifications, Energy 367
  (2026) 142825.
\newblock \href {https://doi.org/10.1016/j.energy.2026.142825}
  {\path{doi:10.1016/j.energy.2026.142825}}.

\bibitem{liu.etal2026LargeLanguage}
D.~Liu, X.~Zhou, Y.~Li, Large language model-driven building energy model
  generation, Energy and Buildings 359 (2026) 117255.
\newblock \href {https://doi.org/10.1016/j.enbuild.2026.117255}
  {\path{doi:10.1016/j.enbuild.2026.117255}}.

\bibitem{zhang.etal2025AutomaticBuilding}
L.~Zhang, V.~Ford, Z.~Chen, J.~Chen, Automatic building energy model
  development and debugging using large language models agentic workflow,
  Energy and Buildings 327 (2025) 115116.
\newblock \href {https://doi.org/10.1016/j.enbuild.2024.115116}
  {\path{doi:10.1016/j.enbuild.2024.115116}}.

\bibitem{lee.yoon2026KnowledgebasedMultiagent}
G.~Lee, S.~Yoon, A knowledge-based multi-agent system using large language
  models for automated building energy modeling and simulation service, Energy
  and Buildings 367 (2026) 117688.
\newblock \href {https://doi.org/10.1016/j.enbuild.2026.117688}
  {\path{doi:10.1016/j.enbuild.2026.117688}}.

\bibitem{jiang.peng2026LargeLanguage}
W.~Jiang, D.~Z. Peng, A large language model-driven automated workflow for
  {CFD} simulations of urban street canyons, Energy and Buildings 366 (2026)
  117694.
\newblock \href {https://doi.org/10.1016/j.enbuild.2026.117694}
  {\path{doi:10.1016/j.enbuild.2026.117694}}.

\bibitem{jiang.chen2027LLMBEMengineerLarge}
G.~Jiang, J.~Chen, {LLM-BEM-engineer}: a large language model-based multi-agent
  {AI} engineer for automated building energy modeling and analysis, Advanced
  Engineering Informatics 77 (2027) 105209.
\newblock \href {https://doi.org/10.1016/j.aei.2026.105209}
  {\path{doi:10.1016/j.aei.2026.105209}}.

\bibitem{evtikhiev.etal2023OutBLEU}
M.~Evtikhiev, E.~Bogomolov, Y.~Sokolov, T.~Bryksin, Out of the {BLEU}: How
  should we assess quality of the code generation models?, Journal of Systems
  and Software 203 (2023) 111741.
\newblock \href {https://doi.org/10.1016/j.jss.2023.111741}
  {\path{doi:10.1016/j.jss.2023.111741}}.

\bibitem{chen.etal2021EvaluatingLarge}
M.~Chen, J.~Tworek, H.~Jun, Q.~Yuan, H.~P. d.~O. Pinto, J.~Kaplan, H.~Edwards,
  Y.~Burda, N.~Joseph, G.~Brockman, et~al., Evaluating large language models
  trained on code (2021).
\newblock \href {http://arxiv.org/abs/2107.03374} {\path{arXiv:2107.03374}},
  \href {https://doi.org/10.48550/arXiv.2107.03374}
  {\path{doi:10.48550/arXiv.2107.03374}}.

\bibitem{jiang.etal2024EPlusLLMLarge}
G.~Jiang, Z.~Ma, L.~Zhang, J.~Chen, {EPlus-LLM}: A large language model-based
  computing platform for automated building energy modeling, Applied Energy 367
  (2024) 123431.
\newblock \href {https://doi.org/10.1016/j.apenergy.2024.123431}
  {\path{doi:10.1016/j.apenergy.2024.123431}}.

\bibitem{jiang.chen2025EfficientFinetuning}
G.~Jiang, J.~Chen, Efficient fine-tuning of large language models for automated
  building energy modeling in complex cases, Automation in Construction 175
  (2025) 106223.
\newblock \href {https://doi.org/10.1016/j.autcon.2025.106223}
  {\path{doi:10.1016/j.autcon.2025.106223}}.

\bibitem{sturmer.etal2026SimulationCode}
J.~M. St{\"u}rmer, J.~Knack, T.~Koch, A.~Weinmann,
  \href{https://arxiv.org/abs/2607.29389}{Simulation code generation for fluid
  systems using large language models: Benchmarking models and prompting
  strategies} (2026).
\newblock \href {http://arxiv.org/abs/2607.29389} {\path{arXiv:2607.29389}}.
\newline\urlprefix\url{https://arxiv.org/abs/2607.29389}

\bibitem{googleGoogleTranslate}
{Google LLC}, {Google Translate}, \url{https://translate.google.com}, accessed
  18 November 2026 (2026).

\bibitem{kwon.etal2023EfficientMemory}
W.~Kwon, Z.~Li, S.~Zhuang, Y.~Sheng, L.~Zheng, C.~H. Yu, J.~Gonzalez, H.~Zhang,
  I.~Stoica, Efficient memory management for large language model serving with
  {PagedAttention}, in: Proceedings of the 29th Symposium on Operating Systems
  Principles, ACM, 2023, pp. 611--626.
\newblock \href {https://doi.org/10.1145/3600006.3613165}
  {\path{doi:10.1145/3600006.3613165}}.

\end{thebibliography}

\ifjournalbuild\else
\clearpage
\begin{center}\Large\bfseries Supplementary material\end{center}

\section*{Supplementary Note S1. Steady state and physical consistency of the reference models}\label{supp:s1}

In the reference set, the residual change of up to 35 K per 1000 s occurs in the rooms of the two topologies that are simulated over 100 s (topologies 1 and 2), whereas the reference models of the other twelve topologies end with every node temperature changing by less than 0.05 K per 1000 s. In topologies 1 and 2, a generated model is simulated over the same simulation horizon as its reference model and the two are compared at L4 at the same instant; the tolerance bands of these topologies were obtained over the same simulation horizon.

In the check of the physical consistency of the reference models (Section 3.1), the energy balance of every mixer closed within 0.01 K, expressed as an equivalent air temperature, and the condensate flow reported by every cooling coil equaled the moisture removed from the air within the rounding of the reported values. Every rotor closed its moisture balance within 0.15\% and its air mass flows within 0.05\%, and its energy balance within 0.3 kW; the largest mismatch occurs in topology 7 and is 2.6\% of the 10.9 kW enthalpy change of its air streams, and the largest relative mismatch is 8\%, 0.08 kW against 1.0 kW, in topology 12. The component-level results are released with the dataset, together with the balances of the rooms, which in the two topologies simulated over 100 s (topologies 1 and 2) are still approaching steady state at the end of the run.

The generated models and the reference model are simulated over the same horizon; in topology 1 the room lies downstream of the monitored supply node, so the supply state does not depend on it, whereas in topology 2 the room air recirculates to the mixer, so the supply-air temperature at 100 s depends on the volume and the initial temperature of the room, which the description does not state. A model of topology 2 built with the library defaults for these two parameters (165 m$^{3}$ and 25\,$^{\circ}$C) instead of the reference values (200 m$^{3}$ and 26\,$^{\circ}$C) has a supply-air temperature 0.43 K below the reference (19.01 against 19.44\,$^{\circ}$C), inside the tolerance band, which extends from 18.93 to 19.94\,$^{\circ}$C.

\section*{Supplementary Note S2. Parameters excluded from the sizing perturbation}\label{supp:s2}

Parameters whose perturbation does not represent design variation were excluded from the perturbation of Section 2.3: absolute temperatures, because a relative perturbation of a temperature changes the operating point rather than the design and produces non-physical variants such as 63\,$^{\circ}$C chilled water; initial conditions, which are numerical rather than physical choices; ambient pressure, a fixed environmental condition; and the adsorption properties of the rotor desiccant, which are fixed by the desiccant material named in the prompt. Parameters that the reference specification does not state keep their defaults from the component block library and were not perturbed either; they include every sizing parameter of the cooling coils in six topologies and the rotor geometry of topology 7.

\section*{Supplementary Note S3. Check of passing specifications against the described operating values}\label{supp:s3}

Every specification that reached L4 was compared statically, without simulation, with the operating values stated in its description: inlet air states and flow rates, coolant flow rates and temperatures, heating-medium temperatures, split ratios, rotor speeds, humidification, and internal heat and moisture loads. A stated value was taken as kept when the specification matched it within 6\% for flow rates, rotor speeds, and heat loads, which the descriptions give as approximate values; within 13\% for moisture loads, which they give to one or two significant figures; within 0.6 K for temperatures; and within 0.02 for relative humidity and split ratios.

\section*{Supplementary Note S4. Screening and translation of the descriptions}\label{supp:s4}

The engineers returned eight descriptions, two each, and these were screened before use for value fidelity, system fidelity, monitored-node declaration, contamination, and format. Six were used. In four of them, sentences outside the system description were deleted, an instruction was rewritten as a statement, or a typing error was corrected, and in two of them the authors of this study added the chilled-water temperature of 5\,$^{\circ}$C, which the table of operating values had omitted. Two were excluded because the air path they described was incomplete or did not match the reference topology, since their cells would measure input error rather than model capability. The screening procedure and criteria are released with the dataset; the texts of the excluded items are not. After the translation into English, the leak scan for framework identifiers was re-run on the English text.

\section*{Supplementary Note S5. Format template and output budget}\label{supp:s5}

The format template is a skeleton of the specification with empty fields and no sizing value, and its uncommented values are placeholders: one air source at 25\,$^{\circ}$C and 50\% relative humidity, and the solver type and maximum step, which are the same as in every reference specification. Its other values are commented examples: an outdoor-air source at 30\,$^{\circ}$C and 70\% relative humidity, a reheat coil with a heating medium at 40\,$^{\circ}$C, and the declaration of a recirculation loop. The same outdoor-air condition appears in the descriptions of topologies 1, 2, 4, 5, 6, and 7, and the same heating-medium temperature of the reheat coil appears in those of topologies 1 and 3. The name of the output-budget parameter in the API request differs across providers, but the value is identical.

\fi

\end{document}